\documentclass[aps,prl,twocolumn,groupedaddress]{revtex4}

\usepackage{graphicx} 
\usepackage{dcolumn}
\usepackage{bm}

\begin{document}
\title{Direct evidence for predominantly phonon-mediated pairing in high-temperature
superconductors} 
\author{Guo-meng Zhao$^{*}$} 
\affiliation{Department of Physics and Astronomy, 
California State University, Los Angeles, CA 90032, USA}

\begin{abstract}
The spectra of the second derivative of tunneling current 
d$^{2}I$/d$V^{2}$ in 
the high-temperature superconductors YBa$_{2}$Cu$_{3}$O$_{7-\delta}$ 
and Bi$_{2}$Sr$_{2}$CaCu$_{2}$O$_{8+\delta}$ 
show clear dip and peak features due to strong coupling to the 
bosonic modes mediating electron pairing. The energy positions of 
nearly all the peaks in $-$d$^{2}I$/d$V^{2}$-like spectra match precisely with those in the phonon density of states obtained by 
inelastic neutron scattering.  The results demonstrate that 
the bosonic modes mediating the electron pairing are 
phonons and that high-temperature superconductivity should arise primarily from strong 
coupling to multiple phonon modes.
 
\end{abstract}
\maketitle 

The identification of phonon anomalies in the tunneling spectra 
\cite{MR} and the observation of an isotope effect \cite{Max} provide 
important and 
crucial clues to the understanding of the microscopic pairing 
mechanism of superconductivity \cite{BCS} in conventional metals. For 
high-transition-temperature (high-$T_{c}$) copper-oxide 
superconductors, 
extensive studies of various unconventional oxygen-isotope effects 
have clearly shown strong electron-phonon interactions and the 
existence of polarons 
\cite{ZhaoAF,ZhaoYBCO,ZhaoLSCO,ZhaoNature97,ZhaoJPCM,Lanzara,SG,HoferPRL,Zhaoreview1,Zhaoreview2,Zhaoisotope,Keller1}. 
Neutron scattering 
\cite{McQueeney}, angle-resolved photoemission (ARPES) 
\cite{Ros}, and Raman scattering \cite{Mis} experiments also provide evidence for the strong electron-phonon 
coupling. However, these 
experiments do not provide direct evidence that pairing is predominantly mediated by 
phonons. Recent ARPES data concerning the effects of electron-boson 
interactions on electron self-energies have led to contradictory 
conclusions.  On one hand, in the diagonal direction of momentum 
space ($\pi$, $\pi$), the observed `kink' features around 70 meV in 
the band 
dispersion of several cuprate compounds appear to provide evidence 
for strong coupling between electrons and the 70 meV Cu-O 
half-breathing mode \cite{Lanzara01}. The coupling to this 70 meV 
phonon mode is 
shown to be primarily responsible for $d$-wave high-$T_{c}$ 
superconductivity \cite{Shen}. On the other hand, Devereaux {\em et al.} propose 
that the 36 meV $B_{1g}$ Cu-O buckling mode rather than the 70 meV mode is 
the main player for $d$-wave high-$T_{c}$ superconductivity based on 
ARPES 
data near the antinodal direction \cite{Devereaux04}. To further 
complicate 
matters, a third group \cite{Gromko} has even provided a completely 
different 
interpretation for the ARPES data in the antinodal direction, that 
is, the kink feature is related to a magnetic resonance mode rather 
than a phonon mode. It is clear that the identity for the bosonic 
modes mediating the electron pairing is inconclusive due to these 
contradictory interpretations.

In order to unambiguously identify the origin of the bosonic modes 
mediating the electron pairing, it is essential to precisely 
determine the energies of the bosonic modes. If the energies of the 
bosonic modes have a one-to-one correspondence to the energies of the 
peak positions in the phonon density of states determined from inelastic 
neutron scattering, one can definitively conclude that phonons 
predominantly mediate the pairing.  For conventional superconductors, 
strong electron-phonon coupling features clearly show up in 
single-particle tunneling spectra \cite{MR,Carb}. The energies  of the 
phonon 
modes coupled to electrons can be precisely determined from tunneling 
spectra. More specifically, the energy positions of the peaks in 
$-$d$^{2}I$/d$V^{2}$, 
measured from the isotropic $s$-wave superconducting gap $\Delta$, 
correspond 
to those of the peaks in the electron-phonon spectral function 
$\alpha{^2}(\omega)$$F(\omega)$ (Refs.~\cite{MR,Carb}). Therefore, if 
the coupling strength $\alpha^{2}(\omega)$ does not have pronounced 
structures, the structures in the phonon density of states $F(\omega)$ 
will have a one-to-one correspondence to those in 
$-$d$^{2}I$/d$V^{2}$.  Such a conventional approach to identify the 
electron-boson coupling features would have been extensively applied 
to high-$T_{c}$ superconductors if the superconducting gap were 
isotropic.  Since the superconducting gap is highly anisotropic in 
high-$T_{c}$ superconductors, it is difficult to reliably determine 
the energies of bosonic modes if tunneling current is not 
directional.  This may explain why the electron-boson coupling structures 
extracted from earlier tunneling spectra are not reproducible among 
different groups \cite{Ved,Shim,Gonnelli}.  In a very recent article 
\cite{Lee} attempting to show an important role of phonons in the 
electron pairing, the authors assign the energy (52 meV) of a peak position  in $+$d$^{2}I$/d$V^{2}$ spectra of 
Bi$_{2}$Sr$_{2}$CaCu$_{2}$O$_{8+\delta}$ to the energy of a phonon 
mode.  Such an assignment is incorrect because the energies of phonon 
modes are equal to the energies of dip positions rather than peak 
positions in $+$d$^{2}I$/d$V^{2}$ (Refs.~\cite{MR,Carb}).

Here we report $Ð$d$^{2}I$/d$V^{2}$-like spectra for the high-$T_{c}$ 
superconductors YBa$_{2}$Cu$_{3}$O$_{7-\delta}$ (YBCO) and 
Bi$_{2}$Sr$_{2}$CaCu$_{2}$O$_{8+\delta}$ (BSCCO). We 
find that the energy positions of nearly all the peaks  in $Ð$d$^{2}I$/d$V^{2}$-like 
spectra match precisely with those of the peaks in the phonon density 
of states obtained by inelastic neutron scattering.  Such excellent 
agreement between tunneling and neutron data can be explained only if 
the tunneling currents in these junctions are highly directional and 
the phonon modes primarily contribute to the electron pairing.

\begin{figure}[htb]
    \includegraphics[height=6.2cm]{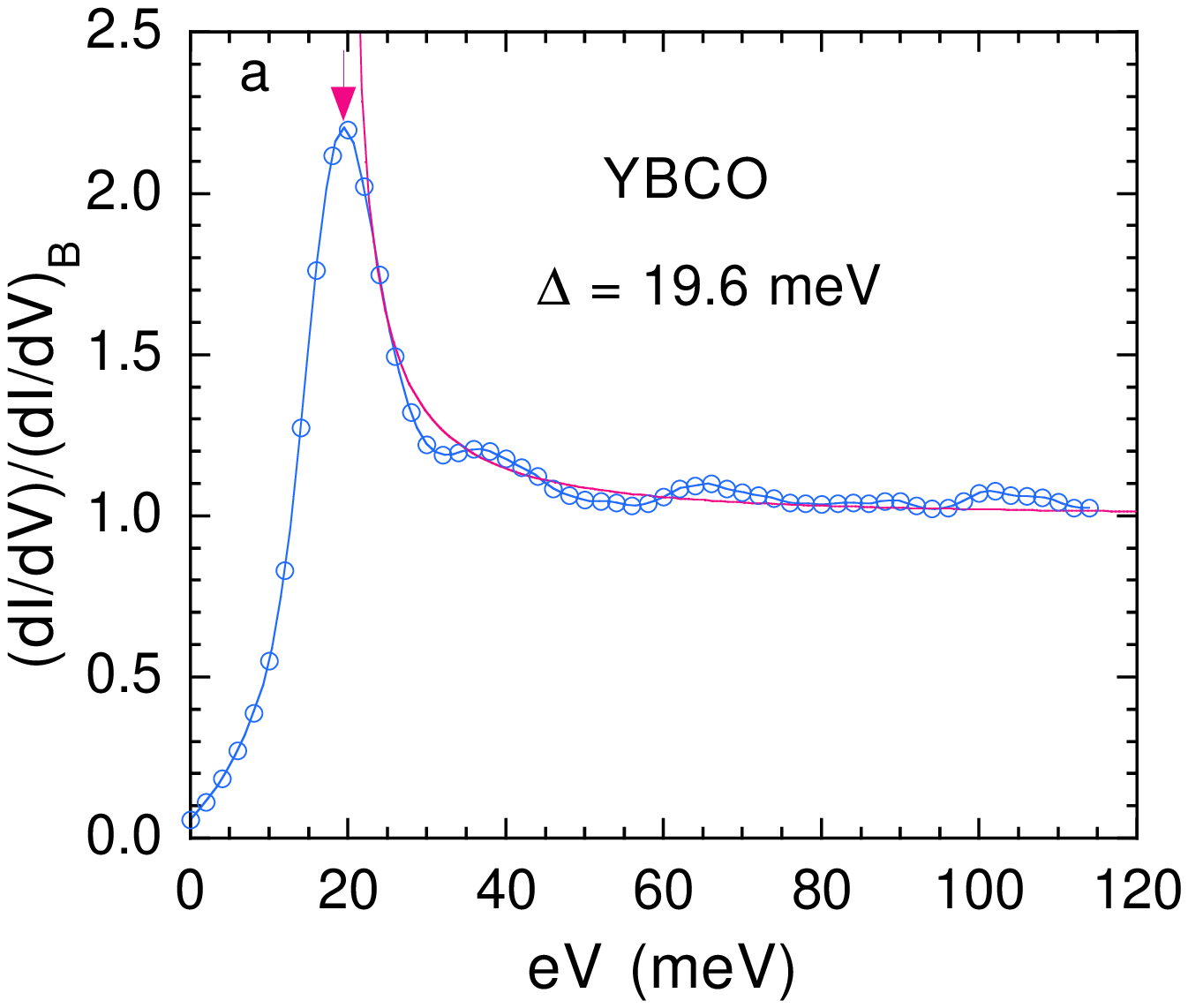}
	 \includegraphics[height=6.2cm]{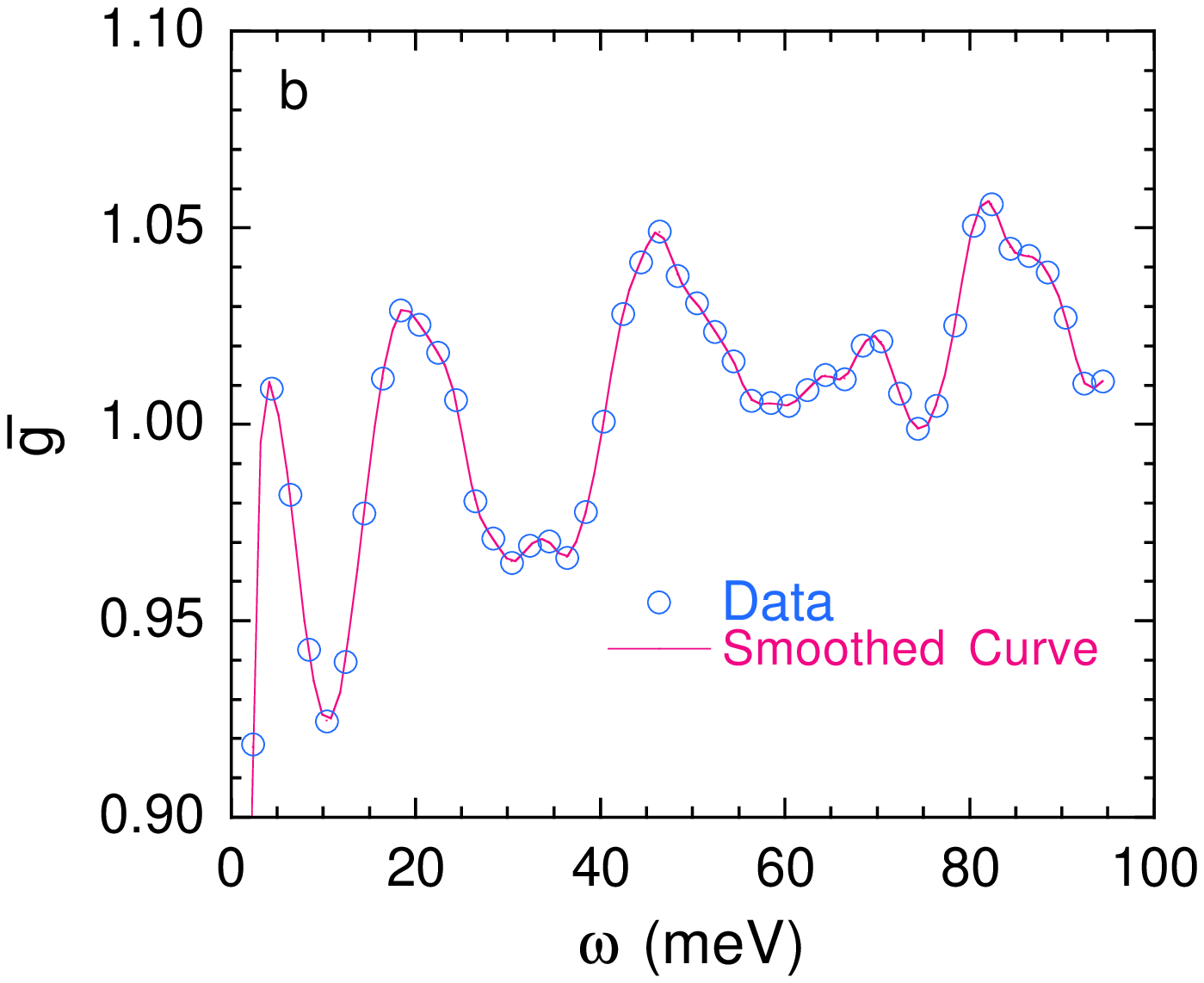}
 \caption[~]{a) Normalized STM conductance $g_{NM}$ = 
 (d$I$/d$V$)/(d$I$/d$V$)$_{B}$ versus energy $E = eV$ for a slightly 
 overdoped YBa$_{2}$Cu$_{3}$O$_{7-\delta}$ (YBCO) with $T_{c}$ = 90 K, 
 where (d$I$/d$V$)$_{B}$ is the background conductance at high biases.  
 The STM conductance spectrum was taken on the (001) crystal face with 
 a Pt-Ir tip at 4.2 K \cite{Wei}.  b) The renormalized conductance 
 $\bar{g}$ = $g_{NM}/N_{BCS}$ versus $\omega = eV-\Delta $ (the energy 
 measured from the gap $\Delta $) for YBa$_{2}$Cu$_{3}$O$_{7-\delta}$.  
 The solid line is a smoothed curve obtained using a cubic-spline 
 interpolation.  }
\end{figure}

Figure~1a shows the normalized scanning tunneling microscopic (STM) conductance 
spectrum $g_{NM}$ = d$I$/d$V$)/(d$I$/d$V$)$_{B}$ at 4.2 K for a 
slightly overdoped YBCO with $T_{c}$ = 90 K, where (d$I$/d$V$)$_{B}$ 
is the background conductance at high biases.  The data are taken from 
Ref.~\cite{Wei} and the normalized conductance is enlarged by a factor 
of 1.06 to ensure unity at high biases.  The spectrum shows negligible 
zero-bias conductance and sharp peak at $eV$ =19.6 meV, suggesting 
that the superconducting gap $\Delta$ is about 19.6 meV.  The solid 
red line is the BCS density of states: $N_{BCS}= 
eV/\sqrt{(eV)^{2}-\Delta^{2}}$, which lies in the middle of the data 
points.  In Fig.~1b, we plot renormalized conductance $\bar{g}$ = 
$g_{NM}/N_{BCS}$ versus $\omega$ = $eV - \Delta$ (the energy measured from 
the gap).  After the renormalization, the structures due to strong 
electron-boson coupling can be clearly seen in the first 
derivative spectrum of tunneling current.  

\begin{figure}[htb]
    \includegraphics[height=6.2cm]{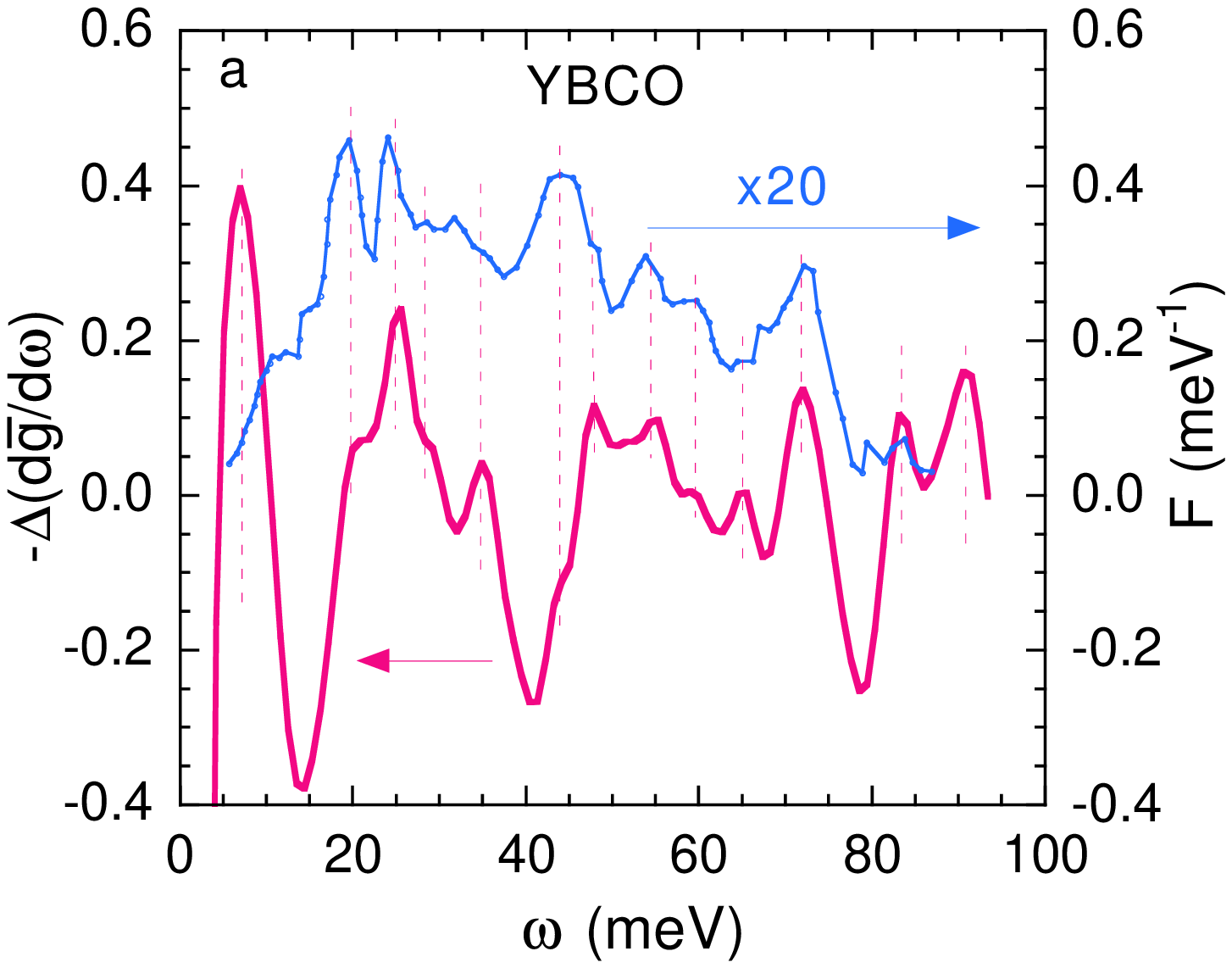}
	 \includegraphics[height=6.2cm]{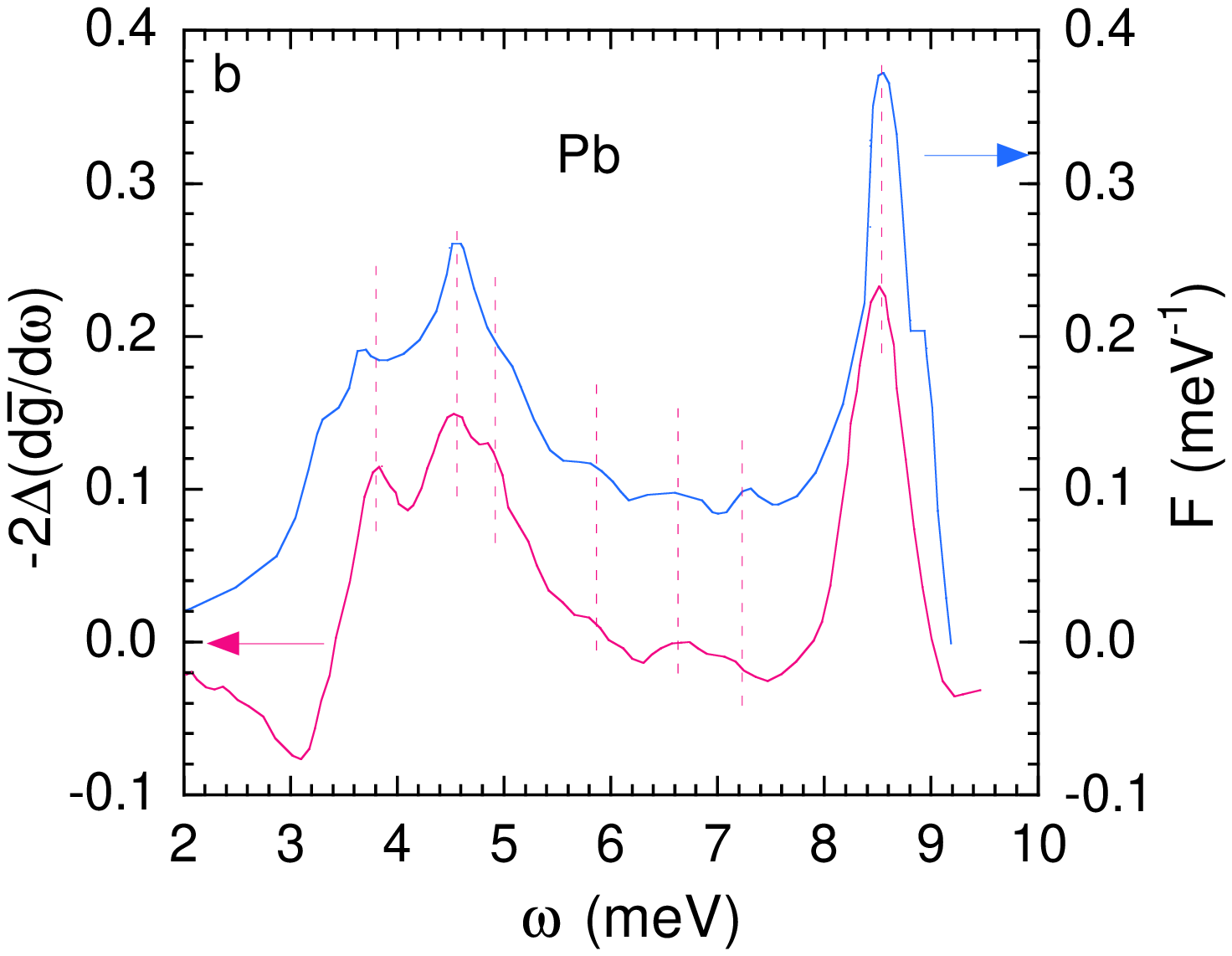}
	\caption[~]{a) $-\Delta$(d$\bar{g}$/d$\omega$) spectrum for the 
YBa$_{2}$Cu$_{3}$O$_{7-\delta}$  crystal together 
with the phonon density of states $F(\omega)$ obtained from inelastic 
neutron scattering \cite{Renker2}.  The left scale is for 
$-\Delta$(d$\bar{g}$/d$\omega$) and the right scale is for 
$F(\omega)$.  The vertical dashed lines mark peak/shoulder 
features in $-\Delta$(d$\bar{g}$/d$\omega$).  b) $-2\Delta$(d$\bar{g}$/d$\omega$) spectrum together 
with the phonon density of states $F(\omega)$ for the 
conventional phonon-mediated superconductor Pb.  Here $\bar{g}$ = 
$N_{s}/(N_{n}N_{BCS})$, $N_{s}$ is the superconducting density of states, and 
$N_{n}$ is the density of states in the normal state.   The left scale is for 
$-2\Delta$(d$\bar{g}$/d$\omega$) and the right scale is for 
$F(\omega)$.  The data of $N_{s}/N_{n}$ are from Ref.~\cite{MR}, and 
the data of $F(\omega)$ are from \cite{Stedman}.  The vertical dashed lines 
mark peak/shoulder features in $-\Delta$(d$\bar{g}$/d$\omega$).  }
\end{figure}

In order to precisely determine the energies of the bosonic modes coupled to electrons, it 
is essential to take the derivative of the renormalized conductance 
$\bar{g}$.  In Fig.~2a, we show the $-\Delta$(d$\bar{g}$/d$\omega$) spectrum 
of the YBCO crystal together with the phonon density of states $F(\omega)$  obtained by inelastic 
neutron scattering \cite{Renker2}.  For comparison,  in Fig.~2b we show 
the $-\Delta$(d$\bar{g}$/d$\omega$) spectrum and $F(\omega)$ for the 
conventional phonon-mediated superconductor Pb.  It 
is apparent that the $-\Delta$(d$\bar{g}$/d$\omega$) spectra for both YBCO 
and Pb show peak, dip, or shoulder features.  A single broad 
peak or shoulder in $F(\omega)$ and $-\Delta$(d$\bar{g}$/d$\omega$) 
can occur when the separations of multiple peak features are too small 
compared to their widths.  For Pb, nearly all the 
peaks/shoulder features in $-\Delta$(d$\bar{g}$/d$\omega$) match precisely with those in
$F(\omega)$, as  expected from the phonon-mediated 
pairing mechanism. However, the peaks at 3.3 meV and 8.9 meV in the phonon density of states correspond to 
the dips in the tunneling spectrum.  This mismatch is due to the fact 
that the coupling strengths for the two modes are much weaker.

If strong coupling to phonon modes also 
happens in high-$T_{c}$ superconductors such as YBCO, then the peak/shoulder 
features in $-\Delta$(d$\bar{g}$/d$\omega$) should line up with the 
peak/shoulder features in the phonon density of states as well.  Indeed,
nearly all 13 peak/shoulder features in 
$-\Delta$(d$\bar{g}$/d$\omega$) of YBCO match precisely with those in 
$F(\omega)$. The strong coupling features at 7.1 meV and 90.8 meV 
in $-\Delta$(d$\bar{g}$/d$\omega$) cannot compare 
with these neutron data since the energy 
positions of 
these features are outside the energy range of the neutron data. 
Nevertheless, the phonon peak at about 6.8 meV is clearly seen in the 
high-resolution neutron data of BSCCO \cite{Renker1} (also see Fig.~3a). The 
strong coupling feature at 83.7 meV does not appear to have a 
correspondence in the neutron data where there is a strong 
localized incoherent peak at 87 meV \cite{AraiYBCO}. However, a clear 
peak feature at about 83.5 meV is seen in the earlier neutron data where 
the localized incoherent peak at 87 meV is not present 
\cite{Renker2}. The feature at 90.8 meV should be the composite phonon 
energy of 7.1 meV and 83.7 meV since the sum of 7.1 meV and 83.7 
meV is equal to 90.8 meV. This is expected from the conventional 
strong-coupling theory \cite{MR,Carb}.  Such excellent 
agreement between neutron and tunneling data provides clear evidence 
that the bosonic modes mediating the electron pairing are phonons 
and that the tunneling current of this junction is highly directional.

It is interesting that the low-energy modes at 7.1 meV, 20.3 meV and 25.5 
meV have much stronger coupling than the higher energy modes and thus 
contribute more to the electron pairing. In contrast, the 
electron-phonon coupling strengths for the 14.6 meV and 31.9 meV phonon modes are 
much weaker because the dip features in 
$-\Delta$(d$\bar{g}$/d$\omega$) occur at these phonon energies.

If the strong electron-phonon coupling features in YBCO are 
intrinsic, they should also show up in other 90 K superconductors. In 
Fig.~3a, we show the $-\Delta$(d$\bar{g}$/d$\omega$) spectrum for 
a BSCCO crystal together with the phonon density of states $F(\omega)$ 
at 296 K.  Here $\bar{g}$ = $N_{s}(\omega)/[N_{n}(\omega)N_{A}(\omega)]$ and $N_{A}(\omega)$ is a 
smoothed curve of $N_{s}(\omega)/N_{n}(\omega)$, which is proportional to $1 + 
0.99\exp (-\omega/18.7) + 0.13\exp [-(\omega -71.3)^{2}/779] $ and 
does not have fine structures.  The data of $N_{s}(\omega)/N_{n}(\omega)$ are 
obtained from Fig.~5a of Ref.~\cite{Gonnelli}.  The 
$N_{s}(\omega)/N_{n}(\omega)$ spectrum has a peak at $eV$ = 25.2 meV (Ref.~\cite{Gonnelli}), 
suggesting $\Delta$ $\simeq$ 25.2 meV. This gap size is smaller than the average gap 
expected for a BSCCO with $T_{c}$= 93 K.  Thus, the spectrum may be 
probing a higher doping region with $T_{c}$ $\simeq$ 89 K in the 
intrinsically inhomogeneous BSCCO.

\begin{figure}[htb]
    \includegraphics[height=6.2cm]{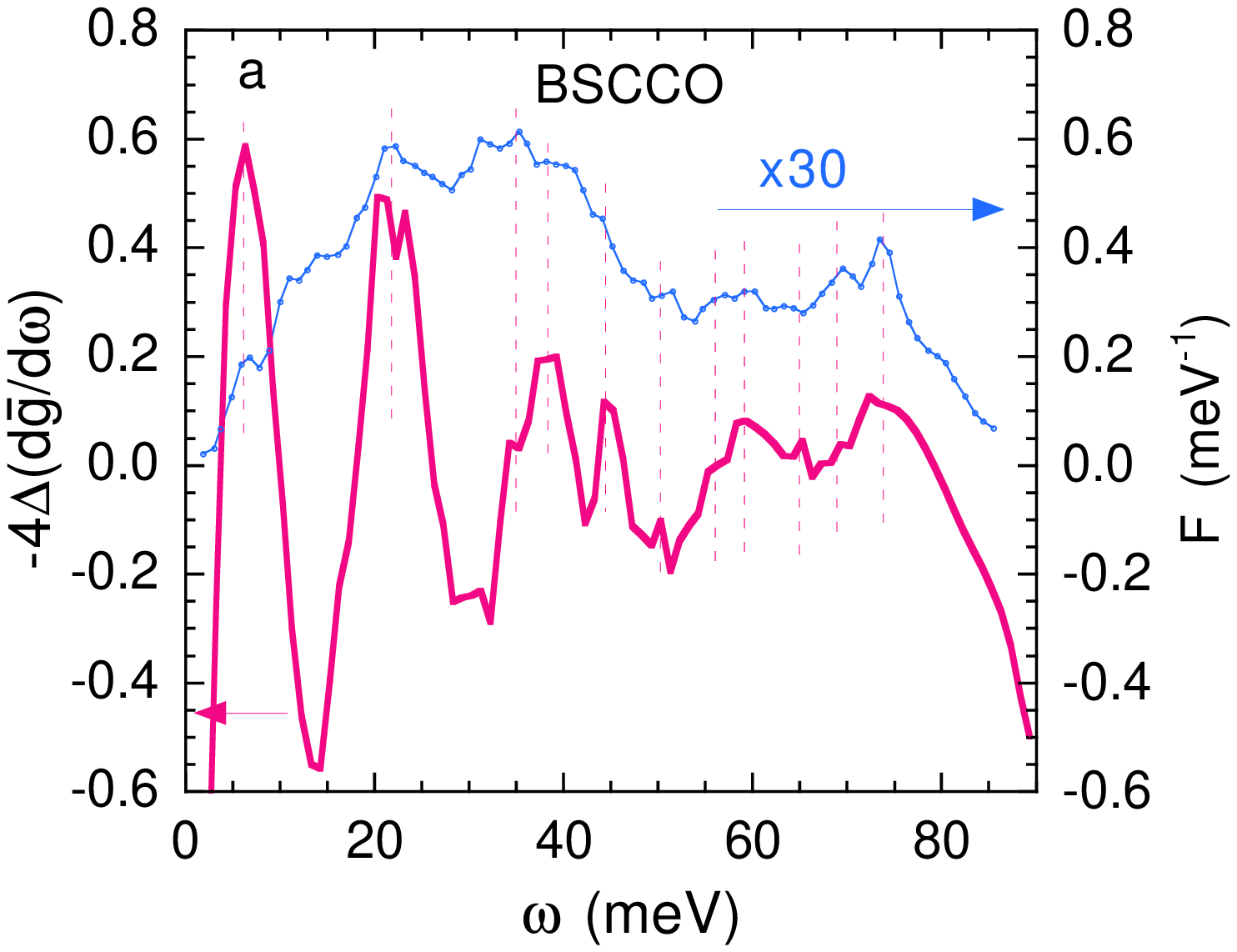}
	 \includegraphics[height=6.2cm]{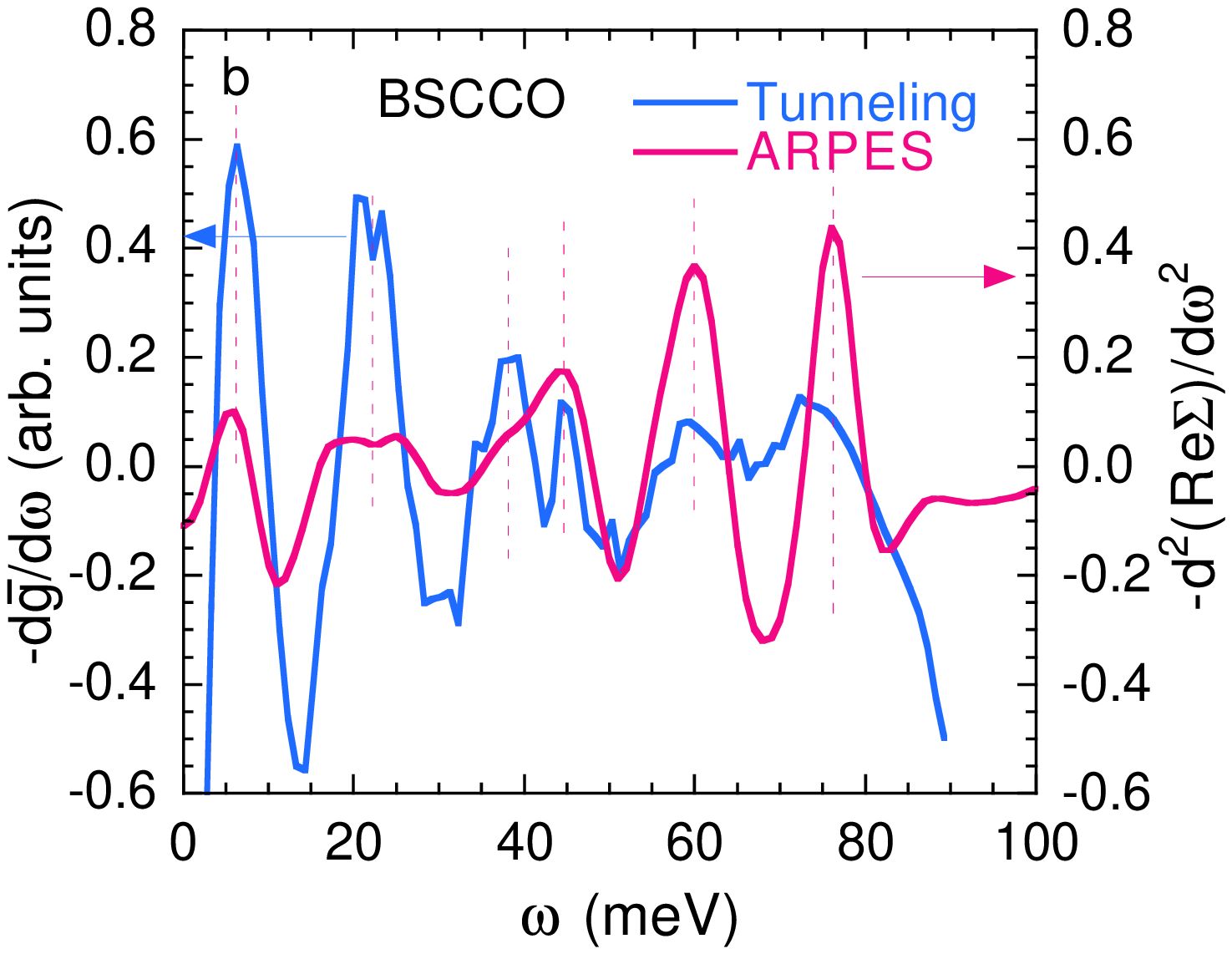}
	\caption[~]{ a) $-\Delta$(d$\bar{g}$/d$\omega$)  spectrum for 
a Bi$_{2}$Sr$_{2}$CaCu$_{2}$O$_{8+\delta}$ crystal with $T_{c}$ = 
93 K and the phonon density of states $F(\omega)$ at 296 K.  The high-resolution 
neutron data of $F(\omega)$ are reproduced from Fig.~1a of Ref.~\cite{Renker1}. The left 
scale is for $-4\Delta$(d$\bar{g}$/d$\omega$) and the right scale is 
for $F(\omega)$.  The vertical dashed lines mark peak/shoulder 
features in $-4\Delta$(d$\bar{g}$/d$\omega$). b) The 
$-$d$^{2}$(Re$\Sigma$)/d$\omega^{2}$ spectrum ($\Sigma$ is electron
self-energy) \cite{Zhao06} and the $-$d$\bar{g}$/d$\omega$ tunneling 
spectrum of BSCCO crystal. The 
$-$d$^{2}$(Re$\Sigma$)/d$\omega^{2}$ spectrum is reproduced from Ref.~\cite{Zhao06}. Vertical dashed lines mark the 
peak positions  in  the $-$d$^{2}$(Re$\Sigma$)/d$\omega^{2}$ spectrum. }
\end{figure}

From Fig.~3a, we can clearly see that nearly all the peak/shoulder 
features in the $-\Delta$(d$\bar{g}$/d$\omega$) spectrum match 
precisely with those in the 
room-temperature phonon density of states. The 
peak feature at 65.0 meV in $-\Delta$(d$\bar{g}$/d$\omega$) is 
slightly off from the peak 
at 64.0 meV in the room-temperature neutron data. It is interesting to
note that two peaks at 14.4 meV and 31.8 meV in the phonon density of states correspond to the dips in the 
tunneling spectrum. For YBCO, this mismatch occurs at 14.6 meV
and 31.9 meV (see
Fig.~2a), which are very close to those (14.4 meV and 31.8 meV) for BSCCO. 
Such quantitative agreement indicates that the 14 meV and 32 meV
phonon modes in both YBCO and BSCCO have much weaker coupling to
electrons.

In Fig.~3b, we compare the strong coupling features revealed in the
tunneling spectrum and in the electron self-energy spectrum of BSCCO.  The spectrum of the 
second derivative of the real part of electron self-energy $-$d$^{2}$(Re$\Sigma$)/d$\omega^{2}$  
is reproduced from Ref.~\cite{Zhao06}. It is striking that the peak features 
in $-$d$^{2}$(Re$\Sigma$)/d$\omega^{2}$ 
match very well with those in $-\Delta$(d$\bar{g}$/d$\omega$). This excellent agreement 
between the 
tunneling and self-energy spectra further demonstrates that the 
observed strong-coupling features in both spectra are intrinsic. 

It is known that any bosonic modes that mediate the electron pairing show up in 
tunneling spectra. The present tunneling spectra of both YBCO and BSCCO clearly 
demonstrate that the bosonic modes below 100 meV, which contribute to the electron
pairing,  are 
only the phonon modes. On the other hand, pairing mechanisms based on strong coupling to a 
magnetic resonance mode predict a pronounced dip feature 
at about 40 meV in the $dI/d\omega$ spectra of 90 K superconductors 
\cite{Abanov}. 
However, the dip feature occurs at about 26 meV, which is nearly independent of doping 
\cite{Lee}. This suggests that the dip feature is not caused by strong 
coupling to the magnetic mode. Moreover, the observed large reduced gap (2$\Delta/k_{B}T_{c}$ $>$ 5) implies 
a large coupling constant ($>$2) and a quite low logarithmic mean boson 
energy $\hbar\omega_{\ln}$ ($<$40 meV) \cite{Carb}. This rules out 
pairing mechanisms based on strong coupling to high-energy magnetic 
and/or charge excitations. Therefore, these tunneling spectra indicate
that high-temperature superconductivity should arise primarily from strong 
coupling to multiple phonon modes.

In summary, we have shown that the energy positions of nearly all the 
peaks in 
$-$d$^{2}I$/d$V^{2}$-like spectra of YBa$_{2}$Cu$_{3}$O$_{7-\delta}$ 
and Bi$_{2}$Sr$_{2}$CaCu$_{2}$O$_{8+\delta}$ match precisely with 
those of the peaks in the phonon density of states obtained by 
inelastic neutron scattering.  The results  demonstrate that 
the bosonic modes mediating the electron pairing are phonons and that 
high-temperature superconductivity should arise primarily from strong 
coupling to multiple phonon modes. The identification of an extended 
$s$-wave gap symmetry in the bulk \cite{Zhaosymmetry} supports the 
phonon-mediated pairing mechanism.
 ~\\
 ~\\
\noindent 
{\bf 
Acknowledgment:} We would like to thank J.  Mann for his critical 
reading and comment.  ~\\
~\\
$^{*}$ gzhao2@calstatela.edu
\bibliographystyle{prsty}

\end{document}